\RequirePackage{fix-cm}
\documentclass[smallextended,refree]{svjour3}       
\smartqed  
\usepackage{amssymb}
\usepackage{latexsym}
\usepackage[dvipdfmx]{graphicx}
\begin{document}

\title{
Fidelity-susceptibility analysis of the 
honeycomb-lattice Ising antiferromagnet
under the imaginary magnetic field
}
\subtitle{}


\author{Yoshihiro Nishiyama
}

\institute{Department of Physics, Faculty of Science,
Okayama University, Okayama 700-8530, Japan}

\date{Received: date / Accepted: date}

\maketitle

\begin{abstract}
The honeycomb-lattice Ising antiferromagnet
subjected to the imaginary magnetic field
$H=i\theta T /2$
with the ``topological'' angle $\theta$
and temperature $T$
was investigated numerically.
In order to treat such a complex-valued statistical weight,
we employed the transfer-matrix method.
As a probe to detect the order-disorder phase transition,
we resort to
an extended version of the fidelity $F$, which makes sense even for such 
a non-hermitian
transfer matrix.
As a preliminary survey,
for an intermediate value of $\theta$,
we investigated
the phase transition via the fidelity susceptibility $\chi_F^{(\theta)}$.
The fidelity susceptibility $\chi_F^{(\theta)}$
exhibits a notable signature for the criticality
as compared to the ordinary quantifiers such as the magnetic susceptibility.
Thereby, we analyze
the end-point singularity of the order-disorder phase boundary
at $\theta=\pi$.
We cast the $\chi_F^{(\theta)}$ data into the crossover-scaling formula
with $\delta \theta = \pi-\theta$ scaled carefully.
Our result for the crossover exponent $\phi$
seems to differ from the mean-field and square-lattice values,
suggesting that 
the lattice structure renders subtle influences as to the multi-criticality at $\theta=\pi$.

\end{abstract}

\section{\label{section1}Introduction}

The concept of fidelity has been developed in the
field of the quantum dynamics
\cite{Uhlmann76,Jozsa94,Peres84,Gorin06}.
The fidelity $F$ is given by the overlap 
$F=|\langle \theta|\theta+\Delta\theta\rangle|$ 
between the ground states, $|\theta\rangle$ and 
$|\theta+\Delta\theta\rangle$,
with the proximate interaction parameters, 
$\theta$ and $\theta+\Delta\theta$, respectively; see Refs. \cite{Vieira10,Gu10,Dutta15}
for a review.
Meanwhile, it turned out that it detects the quantum phase transitions
rather sensitively
\cite{Quan06,Zanardi06,HQZhou08,Yu09,You11,Mukherjee11,Rossini18}.
Actually, the fidelity susceptibility 
$\chi_F=-\frac{1}{N}\partial_{\Delta\theta}^2F|_{\Delta\theta=0}$
($N$: number of lattice points)
exhibits a pronounced signature for the criticality
as compared to the ordinary quantifiers such as the magnetic susceptibility
\cite{Albuquerque10}.
Additionally, the fidelity susceptibility
does not
rely on any presumptions as to the order parameter concerned \cite{Wang15},
and it is less influenced by the finite-size artifacts \cite{Yu09}.
As would be apparent from the definition,
the fidelity $F=|\langle \theta | \theta+\Delta \theta\rangle|$ 
fits the numerical diagonalization method,
which admits the ground-state vector
$|\theta \rangle$ explicitly.
However,
It has to be mentioned
that the fidelity is accessible  via the 
quantum Monte Carlo method
\cite{Albuquerque10,Wang15,Schwandt09,Grandi11}
and
the experimental observations
\cite{Zhang08,Kolodrubetz13,Gu14} as well.

In this paper,
by the agency of the fidelity,
we investigate
the 
honeycomb-lattice Ising antiferromagnet
under the imaginary magnetic field.
To cope with the complex-valued statistical weight,
we employed
the transfer-matrix method \cite{Forcrand18,Zhou08}
through 
resorting to the extended version of the fidelity
\cite{Schwandt09,Sirker10}
\begin{equation}
\label{extended_fidelity}
F= \sqrt{ 
\frac{
 [	{\bf v}_L(\theta+\Delta \theta) \cdot {\bf v}_R(\theta) ][
	{\bf v}_L(\theta) \cdot {\bf v}_R(\theta+\Delta \theta)]
}{
[	{\bf v}_L(\theta) \cdot {\bf v}_R(\theta)    ][
{\bf v}_L(\theta+\Delta \theta) \cdot {\bf v}_R(\theta+\Delta \theta)]
}
} 
 ,
\end{equation}
which makes sense even for such a non-hermitian transfer matrix.
Here,
the right and left eigenvectors
${\bf v}_{R}(\theta)$ and ${\bf v}_{L}(\theta)$ 
satisfy
\begin{equation}
T(\theta){\bf v}_R(\theta) = \lambda_1 {\bf v}_R(\theta)   
\end{equation}
and
\begin{equation}
^{t} {\bf v}_L  (\theta) T(\theta) = \lambda_1 {}^t {\bf v}_L(\theta)    ,
,
\end{equation}
respectively,
with the maximal eigenvalue 
$\lambda_1$ 
of the transfer matrix $T(\theta)$.
Notably,
the expression (\ref{extended_fidelity}) reduces to the aforementioned one
$|\langle \theta | \theta +\Delta \theta \rangle |$, provided that the matrix $T(\theta)$ is hermitian,
and the vectors ${\bf v}_{R,L}$ are adjoint.
In the present work,
the matrix $T(\theta)$ is non-symmetric and complex-valued, and it is by no means hermitian.
So far, the real-non-symmetric \cite{Sirker10}
and complex-valued-symmetric \cite{Forcrand18,Nishiyama20} transfer matrices
were treated
for the quantum $XXZ$ chain
and square-lattice Ising antiferromagnet under the imaginary field, respectively.
In this paper, we demonstrate that the complex-valued non-symmetric $T$,
namely, the honeycomb-lattice case, is also
tractable with the simulation scheme.
It is anticipated that
the criticality of the honeycomb-lattice model is not identical to that of the square-lattice model,
because
the former possesses the duality
\cite{Imaoka96,Suzuki90,Lin88} at a finely-tuned value of the imaginary magnetic field.

To be specific,
we present the
Hamiltonian for the honeycomb-lattice
Ising antiferromagnet subjected to the imaginary magnetic field
\begin{equation}
\label{Hamiltonian}
{\cal H}
=J \sum_{\langle ij \rangle} S_i S_j -H\sum_i^N S_i
.
\end{equation}
Here, the Ising spin $S_i=\pm 1$
is placed at each honeycomb-lattice point $i=1,2,\dots,N$,
and the summation $\sum_{\langle ij\rangle}$
runs over all possible nearest-neighbor pairs $\langle ij \rangle$.
Hereafter,
the antiferromagnetic coupling constant $J$
is regarded as the unit of energy $J=1$.
The magnetic field $H$ takes a pure imaginary value
\begin{equation}
H=i\theta T/2
 ,
\end{equation}
with the ``topological'' angle $\theta$, and temperature $T$,
and likewise, the reduced coupling constant $K=J/T$ is introduced.

\begin{figure}
\includegraphics{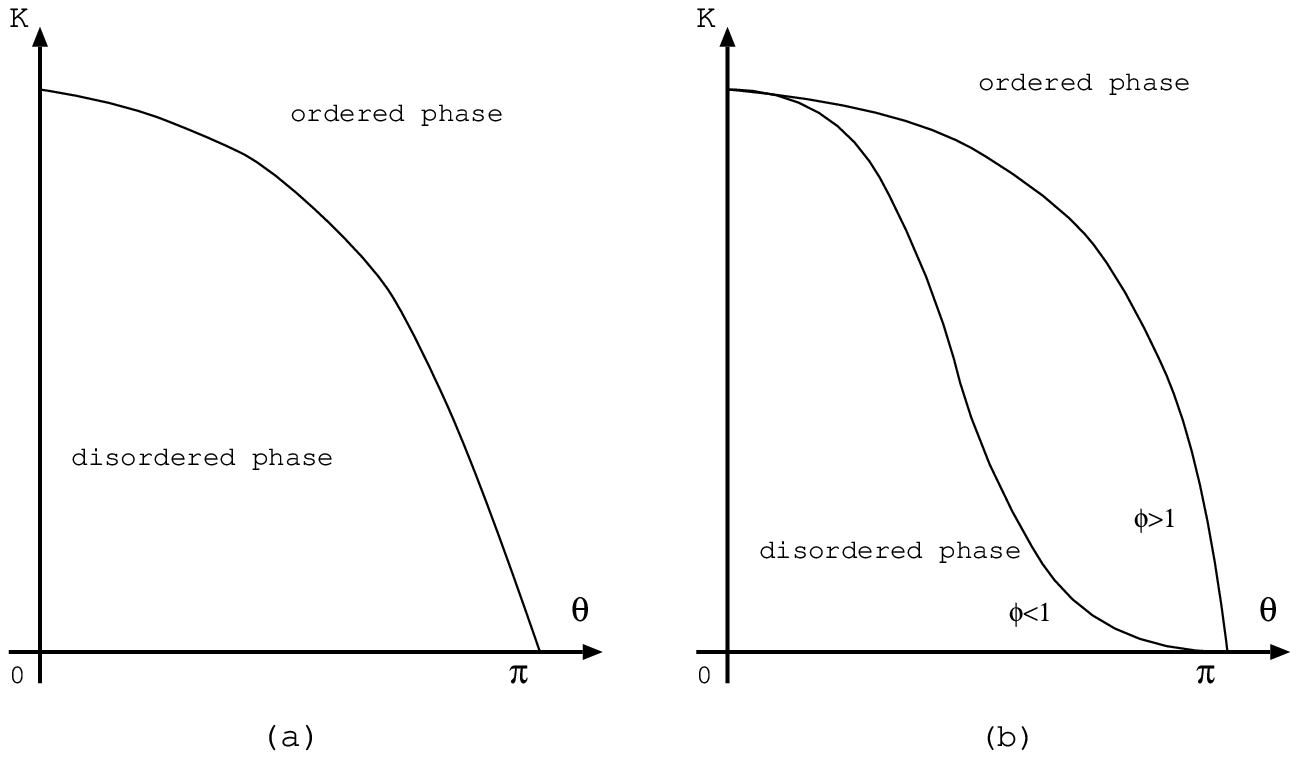}
\caption{
(a)
A schematic $\theta$-$K$
phase diagram is presented for the honeycomb-lattice
Ising antiferromagnet (\ref{Hamiltonian})
under the imaginary magnetic field $H=i\theta T/2$;
here, the reduced coupling constant $K$ is given by $K=J/T$.
The overall feature should resemble the
square-lattice case \cite{Matveev08,Azcoiti17};
a notable point is that
unlike the ferromagnetic counterpart \cite{Lee52},
the order-disorder phase boundary extends into
the $\theta>0$ regime.
The phase boundary ends up at $\theta=\pi$, where
a number of rigorous results are available 
\cite{Imaoka96,Suzuki90,Lin88}.
(b) 
The end-point behavior of the phase boundary 
around $\theta\approx \pi$
may depend on the lattice structure undertaken.
The mean-field analysis \cite{Azcoiti11} suggests that the phase boundary
exhibits a convex curvature
characterized \cite{Riedel69,Pfeuty74}
by the small crossover exponent $\phi=1/2(<1)$.
On the one hand,
For the square-lattice model, 
a concave curvature $\phi>1$
was suggested via the 
partition-function-zeros \cite{Matveev08}
and series-expansion
\cite{Azcoiti17} analyses; see 
Ref. \cite{Nishiyama20} as well.
Then, there arises a problem how the honeycomb-lattice structure 
influences the crossover exponent $\phi$.
}
\label{figure1}       
\end{figure}

A schematic drawing of the phase diagram 
\cite{Imaoka96,Suzuki90,Lin88,Kim10,Matveev96,Matveev96b}
is presented in 
Fig. \ref{figure1} (a).
The order-disorder phase boundary extends into the 
finite-$\theta$ regime,
and eventually, the phase boundary terminates at $\theta=\pi$.
So far, the model (\ref{Hamiltonian})
has been investigated by means of
the partition-function zeros  
\cite{Kim10}, albeit with an emphasis on the {\em real}-$H$-driven criticality.
Additionally, 
rigorous information in terms of the duality theory
\cite{Imaoka96,Suzuki90,Lin88,Matveev96,Matveev96b} is available at $\theta=\pi$.
In fairness, it has to be mentioned that
the {\em square}-lattice counterpart
has been investigated
with the partition-function-zeros \cite{Matveev08},
series-expansion \cite{Azcoiti17}, and
exact-diagonalization \cite{Nishiyama20} methods
in order to surmount the severe sign problem due to 
the imaginary magnetic field \cite{Azcoiti11}.
In particular,
the series expansion has played a significant role;
actually, the above-mentioned works, Ref. 
\cite{Matveev96} and
\cite{Azcoiti17},
made use of the low-temperature and cumulant expansions, respectively,
elucidating the underlying criticality  rather systematically.
The phase diagram, Fig. \ref{figure1} (a),
should resemble that of the square-lattice model in overall characters.
In contrast, 
the ferromagnetic counterpart
does exhibit 
no phase transition 
in the finite-$\theta$ domain
according to the celebrated theory of the
Lee-Yang zeros \cite{Lee52}.

Then, there arises a problem how the phase boundary
ends up at $\theta=\pi$; see Fig. \ref{figure1} (b).
The mean-field result \cite{Azcoiti11} 
shows that the phase boundary is curved convexly around $\theta \approx \pi$,
characterized 
\cite{Riedel69,Pfeuty74}
by the crossover exponent $\phi (=1/2) <1$.
On the contrary,
the above mentioned numerical results \cite{Matveev08,Azcoiti17,Nishiyama20}
for the square-lattice model
indicate a concavely-curved phase boundary with
$\phi >1$.
Because the honeycomb-lattice antiferromagnet
at $\theta=\pi$ 
is under the reign of the duality theory
\cite{Imaoka96,Suzuki90,Lin88,Matveev96}, 
the end-point singularity $\phi$ should reflect
its peculiar characters.
The aim of this paper is to explore $\phi$ quantitatively
by casting the fidelity-susceptibility 
data into the crossover-scaling theory
\cite{Riedel69,Pfeuty74};
a key ingredient is that
the probe $\chi_F^{(\theta)}$ exhibits a notable singularity
at the end-point $\theta=\pi$.

The rest of this paper is organized as follows.
In Sec. \ref{section2},
we present
the numerical results.
The 
transfer-matrix scheme is outlined as well.
In Sec. \ref{section3}, we address the summary and discussions.

\section{\label{section2}Numerical results}

\begin{figure}
\includegraphics{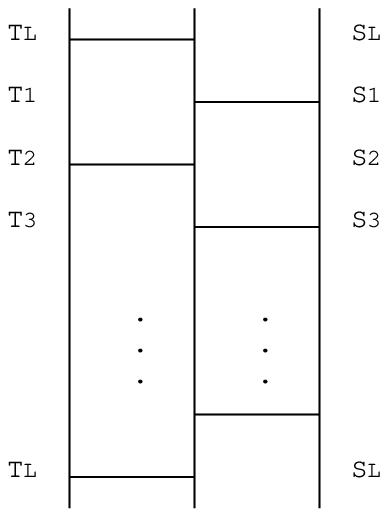}
\caption{
A unit of the transfer-matrix slice
for the honeycomb-lattice Ising antiferromagnet (\ref{Hamiltonian})
is shown.
The row-to-row statistical weight
between the spin configurations, $S_{1,2,\dots,L}$ and $T_{1,2,\dots,L}$,
yields the transfer matrix element $T_{\{S_i\},\{T_i\}}$.
The transfer matrix $T$ is not symmetric as would be apparent from
the drawing.
The periodic-boundary condition
such as $S_{L+1}=S_1$ and $T_{L+1}=T_1$
is imposed.
}
\label{figure2}       
\end{figure}

In this section, we present the numerical results for 
the honeycomb-lattice Ising antiferromagnet
under the imaginary magnetic field (\ref{Hamiltonian}).
In order to cope with the complex-valued statistical weight,
we employed the transfer-matrix method 
as developed in Ref. \cite{Forcrand18},
where 
the authors investigated
the
{\em square}-lattice version rather in detail;
our simulation scheme owes to this development.
In Fig. \ref{figure2},
we present 
a unit of the transfer-matrix slice for the honeycomb-lattice model.
The row-to-row
statistical weight between the spin arrangements,
$S_i$ and $T_i$ ($i=1,2,\dots,L$),
yields the transfer-matrix element, $T_{\{S_i \} , \{T_i \} }$.
Here, we implemented the periodic-boundary condition
such as $S_{L+1}=S_1$ and $T_{L+1}=T_1$.
As would be apparent from Fig. \ref{figure2},
the transfer matrix is not symmetric, 
$T_{\{S_i\} , \{T_i \}} \ne T_{\{T_i\} , \{S_i \}}$,
in contrast to that of the square-lattice case \cite{Forcrand18}.
Correspondingly, in the present case,
the left and right eigenvectors, ${\bf v}_L$ and ${\bf v}_R$,
are neither identical
nor adjoint, and the extention of the fidelity (\ref{extended_fidelity})
now becomes essential.

Provided that the fidelity $F$ 
(\ref{extended_fidelity}) is at hand,
we are able to
evaluate the fidelity susceptibility
\begin{equation}
\label{fidelity_susceptibility}
\chi_F^{(\theta)} = - \frac{1}{L} \partial_{\Delta \theta}^2 F |_{\Delta \theta=0}
.
\end{equation}
According to the scaling theory \cite{Albuquerque10},
the fidelity susceptibility (\ref{fidelity_susceptibility})
exhibits an enhanced 
singularity
as compared to the ordinary quantifiers such as the magnetic susceptibility.
Note that for the two-dimensional Ising antiferromagnet,
both
specific heat and {\em uniform} susceptibility exhibit weak (logarithmic) singularities
at the N\'eel temperature.
Hence, it is significant to search for alternative quantifiers
so as to detect the signature for the criticality.

\subsection{\label{section2_1}
Finite-size-scaling analysis of the fidelity susceptibility $\chi_F^{(\theta)}$ 
(\ref{fidelity_susceptibility})
with the fixed $\theta=2.22$}

In this section,
via the fidelity susceptibility $\chi_F^{(\theta)}$ (\ref{fidelity_susceptibility}),
we investigate the order-disorder phase transition 
of the honeycomb-lattice antiferromagnet with an intermediate value of the imaginary magnetic field
$\theta=2.22$.
At this point $\theta=2.22$,
a preceding simulation result \cite{Kim10} is available.

\begin{figure}
\includegraphics{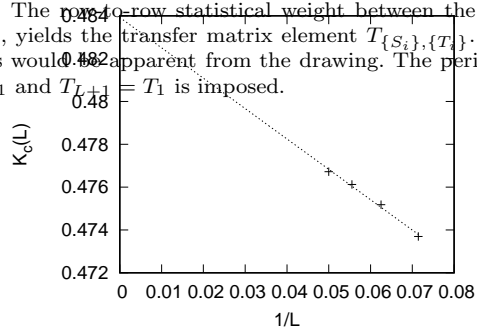}
\caption{
The approximate critical point $K_c(L)$ 
(\ref{approximate_critical_point})
is plotted for $1/L$ with the fixed imaginary magnetic field $\theta=2.22$.
The least-squares fit to the data yields an estimate 
$K_c=0.48392(55)$
in the thermodynamic limit $L\to\infty$.
A possible systematic error is considered in the text.
}
\label{figure3}       
\end{figure}

In Fig. \ref{figure3},
we present the approximate critical point 
$K_c(L)$
for $1/L$
with the fixed $\theta=2.22$ and various system sizes $L=14,16,\dots,20$.
Here, the approximate critical point $K_c(L)$
denotes the
maximal point of the fidelity susceptibility
\begin{equation}
	\label{approximate_critical_point}
\partial_K \chi_F^{(\theta)}|_{K=K_c(L)}=0 ,
\end{equation}
for each $L$.
The least-squares fit to these data yields an estimate
$K_c=0.48392(55)$ in the thermodynamic limit $L \to \infty$.
As a reference, we carried out the similar extrapolation scheme with the abscissa scale replaced
with $1/L^2$, and we arrived at an alternative estimate $K_c=0.47969(12)$.
The deviation between them $\approx 0.004$ dominates
the least-squares-fitting error $\approx 0.00055$.
Hence,
regarding the former $0.004$ as a possible systematic error,
we estimate the critical point as 
\begin{equation}
\label{critical_point}
K_c=0.484(4)
.
\end{equation}

The estimate (\ref{critical_point})
is to be compared with the preceeding result 
$K_c|_{\theta=2.22}=0.458 \dots$ \cite{Kim10};
see Table \ref{table1}.
In Ref. \cite{Kim10}, 
the distribution of the partition-function zeros was explored
for the honeycomb-lattice antiferromagnet,
albeit with an emphasis on the 
{\em real}-$H$-driven phase transition;
afterward, we explain how the transition point was extracted from their
simulation result.
As presented in Table \ref{table1},
our result $K_c|_{\theta=2.22}=0.484(4)$
[Eq. (\ref{critical_point})]
is comparable to this elaborated pioneering study,
$K_c|_{\theta=2.22}=0.458$ \cite{Kim10}.
Such a feature validates the use of
the $\chi_F^{(\theta)}$-mediated simulation scheme
even for
the case of 
non-symmetric complex-valued transfer matrix of the honeycomb-lattice model
(\ref{Hamiltonian}).

\begin{figure}
\includegraphics{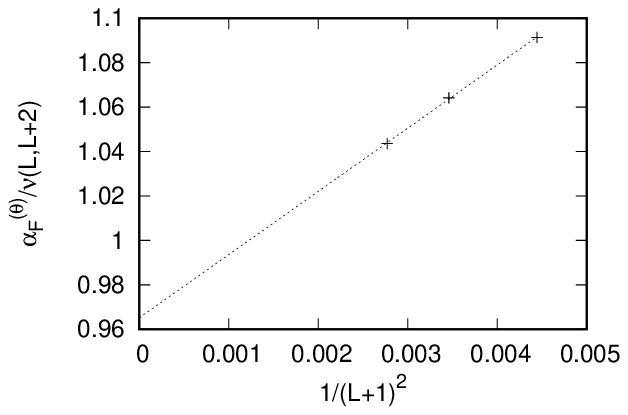}
\caption{
The approximate critical exponent
$\alpha_F^{(\theta)}/\nu(L,L+2)$ 
(\ref{approximate_critical_exponent}) 
is plotted for $1/(L+1)^2$ with the fixed imaginary magnetic field $\theta=2.22$.
The least-squares fit to the data yields an estimate 
$\alpha_F^{(\theta)} / \nu =0.9652(22)$
in the thermodynamic limit $L\to\infty$.
A possible systematic error is considered in the text.
}
\label{figure4}       
\end{figure}

We then turn to the analysis of the criticality,
namely, $\chi_F^{(\theta)}$'s scaling dimension $\alpha_F^{(\theta)}/\nu$
\cite{Albuquerque10}.
Here, the index $\alpha_F^{(\theta)}$ 
($\nu$) denotes the fidelity-susceptibility
(correlation-length) critical exponent such as
$\chi_F^{(\theta)} \sim |K-K_c|^{-\alpha_F^{(\theta)}}$
($\xi \sim |K-K_c|^{-\nu}$).
In Fig. \ref{figure4},
the approximate critical exponent 
$\alpha_F^{(\theta)}/\nu(L,L+2)$ is plotted
for $1/(L+1)^2$ with the fixed $\theta=2.22$
and various system sizes $L=14,16,18$.
Here,
the approximate critical exponent 
is given by the logarithmic derivative of $\ln \chi_F^{(\theta)}(L)$
\begin{equation}
\label{approximate_critical_exponent}
\frac{\alpha_F^{(\theta)} } {\nu}(L,L')
=\frac{\ln (\chi_F^{(\theta)}(L)|_{K_c(L)}   /
\chi_F^{(\theta)}(L')|_{K_c(L')}  )}
    {\ln(L/L')}
,
\end{equation}
for a pair of system sizes $(L,L')$.
The least-squares fit to these data yields an estimate 
$\alpha_F^{(\theta)}/\nu=0.9652(22)$
in the thermodynamic limit $L\to\infty$.
Alternatively, we arrive at 
$\alpha_F^{(\theta)}/\nu=0.9986(38)$ with the abscissa scale replaced with $1/(L+1)^3$.
The deviation between them $\approx 0.03$ dominates the 
least-squares-fitting error $\approx 0.0022$.
Hence, considering the former as  a possible systematic error,
we estimate the critical exponent as
\begin{equation}
\label{critical_exponent}
\alpha_F^{(\theta)}/\nu = 0.97(3) 
  .
\end{equation}

According to the scaling theory \cite{Albuquerque10},
the scaling relation 
\begin{equation}
\label{scaling_relation}
\alpha_F^{(\theta)} / \nu = \gamma_{af} / \nu +1
,
\end{equation}
holds
with the magnetic-susceptibility critical exponent for the antiferromagnet
$\gamma_{af}$.
Putting our result 
$\alpha_F^{(\theta)}/\nu =0.97(3)$
[Eq. (\ref{critical_exponent})]
into this scaling relation (\ref{scaling_relation}), we obtain 
\begin{equation}
\label{critical_exponent_gamma}
\gamma_{af}/\nu = -0.03(3) .
\end{equation}
This result
indicates that the phase transition 
belongs to the two-dimensional-Ising universality class,
$\gamma_{af}=0$ (logarithmic) 
\cite{Fisher60,Kaufman87}.

A few remarks are in order.
First,
we stress that $\chi_F^{(\theta)}$'s scaling dimension, 
$\alpha_F^{(\theta)}/\nu=0.97(3)$
[Eq. (\ref{critical_exponent})],
is larger than that of the magnetic susceptibility,
$\gamma_{af}/\nu=-0.03(3)$
[Eq. (\ref{critical_exponent_gamma})].
Therefore,
the fidelity susceptibility $\chi_F^{(\theta)}$
admits a pronounced signature 
for the criticality as compared to the ordinary quantifiers such as
the magnetic susceptibility.
Such a feature is significant for the two-dimensional Ising antiferromagnet,
where both specific heat and magnetic susceptibility exhibit weak (logarithmic) singularities
at the N\'eel temperature.
Last, we explain
how the critical point $K_c|_{\theta=2.22}=0.458 \dots $     
was extracted from Fig. 2 (d) of Ref. \cite{Kim10}.
In Ref. \cite{Kim10}, the partition-function zeros were calculated for the
complex domain of
$x=\exp (i\theta)$ with generic
$\theta \in \mathbb{C}$; here,
the reduced coupling constant $K$ is fixed to $0.4=\exp(-2K)$.
In Fig. 2  (d) of Ref. \cite{Kim10},
the accumulation of zeros forms a branch, which is about to touch the unit circle $x = \exp(i\theta)$
($\theta \in \mathbb{R}$).
Such a feature 
indicates that
the $\theta(\in \mathbb{R})$-driven phase transition
occurs at this crossing point.
More specifically,
we read off a couple of partition-function zeros
$x_1=-0.5014 + 0.4833 i$ and
$x_2=-0.4644 + 0.3733 i$,
and found that
the line defined by these points
crosses the unit circle at $\theta=2.22\dots$.   
The above data are tabulated 
in Table \ref{table1}.
Nonetheless, we stress
that the mechanism behind the phase transition differs 
from that of the ferromagnetic case \cite{Lee52}, as noted in Ref. \cite{Kim10}.
In the ferromagnetic case, the partition-function zeros simply forms
a unit circle.
Neither
extra branch  nor mutual crossing occurs, and
no $\theta$-driven phase transition takes place at all.

\subsection{\label{section2_2}
Scaling plot for $\chi_F^{(\theta)}$  with the fixed $\theta=2.22$}

In this section, in order to check the validity of the scaling analyses in Sec. \ref{section2_1},
we present $\chi_F^{(\theta)}$'s scaling plot, which also
sets a basis of the 
subsequent crossover-scaling analyses.
The fidelity susceptibility obeys the scaling formula \cite{Albuquerque10}
\begin{equation}
\label{scaling_formula}
\chi_F^{(\theta)} = L^x f \left( (K-K_c)L^{1/\nu} \right)
  ,
\end{equation}
with $\chi_F^{(\theta)}$'s scaling dimension $x=\alpha_F^{(\theta)}/\nu$
and a non-universal scaling function $f$.

\begin{figure}
\includegraphics{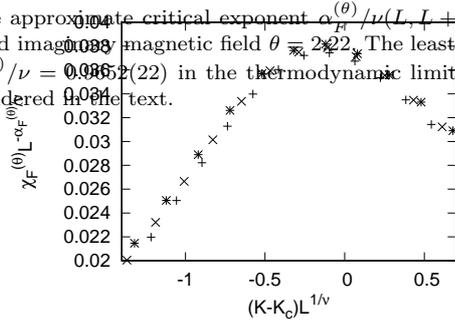}
\caption{
The scaling plot, 
$(K-K_c)L^{1/\nu}$-$\chi_F^{(\theta)}L^{-\alpha_F^{(\theta)}/\nu}$,
is presented
with $K_c=0.484$ 
[Eq. (\ref{critical_point})],
$\alpha_F^{(\theta)}/\nu=0.97$ 
[Eq. (\ref{critical_exponent})],
and $\nu=1$
(2D-Ising universality)
for various system sizes,
($+$) $L=16$,
($\times$) $18$,
and
($*$) $20$;
see the scaling formula
(\ref{scaling_formula}).
The scaled data seem to fall into the scaling curve satisfactorily,
validating the scaling analyses in Sec. \ref{section2_1}
and the proposition $\nu =1$.
}
\label{figure5}       
\end{figure}

In Fig. \ref{figure5}, we present the scaling plot,
$(K-K_c)L^{1/\nu}$-$\chi_F^{(\theta)}L^{-\alpha_F^{(\theta)}/\nu}$,
for various system sizes
($+$) $L=16$,
($\times$) $18$,
and
($*$) $20$ with the fixed $\theta=2.22$.
Here, we made a proposition  $\nu=1$ (two-dimensional-Ising universality),
and the other scaling parameters are set to
$K_c=0.484$ [Eq. (\ref{critical_point})],
and
$\alpha_F^{(\theta)}/\nu=0.97$ [Eq. (\ref{critical_exponent})].
The scaled data in Fig. \ref{figure5}
collapse into the scaling function $f$ satisfactorily,
validating the scaling analyses in Sec. \ref{section2_1}
as well as the proposition $\nu=1$.
Hence, 
recollecting $\gamma_{af}/\nu= - 0.03(3)$ [Eq. (\ref{critical_exponent_gamma})],
we confirm that
the order-disorder phase transition indeed
belongs to the two-dimensional-Ising universality class.

The scaling plot, Fig. \ref{figure5}, indicates that the fidelity susceptibility
is less affected by the finite-size artifact \cite{Yu09}.
Such a feature is favorable for the 
exact diagonalization method,
with 
which the tractable system size is rather restricted.
Encouraged by this observation, we proceed to examine 
how the order-disorder phase boundary terminates
at the extremum point $\theta=\pi$.

\subsection{\label{section2_3}
Crossover-scaling plot for $\chi_F^{(\theta)}$ 
around $\theta \approx \pi$}

In the above section, based on the scaling formula (\ref{scaling_formula}),
we confirmed that the order-disorder-phase-transition branch belongs to the two-dimensional
Ising universality class. 
In this section, by the agency of  $\chi_F^{(\theta)}$,
we further explore the end-point singularity of the phase boundary toward $\theta = \pi$.
For that purpose, introducing yet another controllable parameter
$\delta \theta = \pi -\theta$
and the accompanying crossover exponent $\phi$,
we consider
the crossover-scaling formula \cite{Riedel69,Pfeuty74}
\begin{equation}
\label{crossover_scaling_formula}
	\chi_F^{(\theta)} = L^{\dot{x}}  g \left(
\left( K-K_c(\theta) \right)
	      L^{1/\dot{\nu}},
	\delta \theta L^{\phi/\dot{\nu}} 
\right)
 ,
\end{equation}
with the
$\theta$-dependent
critical point $K_c(\theta)$, and a non-universal scaling function $g$.
Here, the indices, $\dot{x}$ and $\dot{\nu}$,
are $\chi_F^{(\theta)}$'s scaling dimension and correlation-length critical exponent,
respectively, right at $\delta \theta=0$.
As in Eq. (\ref{scaling_formula}), the index
$\dot{x}$ satisfies $\dot{x}=\dot{\alpha}_F^{(\theta)}/\dot{\nu}$
\cite{Albuquerque10}
with the fidelity-susceptibility critical exponent $\dot{\alpha}_F^{(\theta)}$
at $\delta \theta=0$.

As explained in Sec. \ref{section1},
the crossover exponent $\phi$ describes the shape of 
the phase boundary 
as $K_c \sim \delta \theta^{1/\phi}$ \cite{Riedel69,Pfeuty74}.
Hereafter, 
the crossover exponent $\phi$ is considered as an adjustable parameter,
and the other indices, 
$\dot{x}$, $\dot{\alpha}_F^{(\theta)}$,
and $\dot{\nu}$,
are fixed in prior to the scaling analyses as follows.
According to the duality theory \cite{Imaoka96}, 
the hexagonal-lattice Ising antiferromagnet at $\theta=\pi$ 
reduces to the
triangular-lattice antiferromagnet,
and
the uniform-susceptibility and correlation-length exponents
are given by $\dot{\gamma}_{af}=3/2$ and $\dot{\nu}=1$, respectively \cite{Horiguchi92}.
Notably enough,
through the duality,
the frustrated (non-bipartite lattice)  antiferromagnet comes out
from the seemingly non-frustrated magnet, albeit with the imaginary magnetic field mediated.
This is a peculiarity of the imaginary-field magnet, and such a character would not
be captured properly by the mean-field treatment.
These indices together with the relation $\dot{\alpha}_F^{(\theta)}/\dot{\nu}=\dot{\gamma}_{af}/\dot{\nu}+1$
\cite{Albuquerque10}
(see Eq. (\ref{scaling_relation})) immediately
admit $\dot{x}(=\dot{\alpha}_F^{(\theta)}/\dot{\nu})=5/2$,
which now completes the prerequisite for the crossover-scaling analysis.

\begin{figure}
\includegraphics{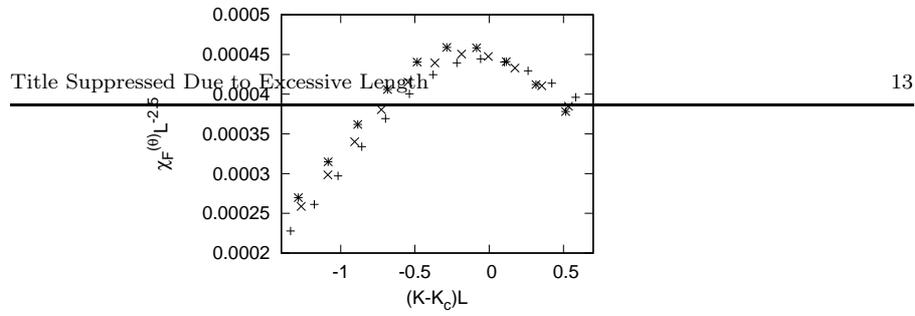}
\caption{
The crossover-scaling plot,
$(K-K_c(\theta))L$-$\chi_F^{(\theta)}L^{-2.5}$,
is presented 
with the
second argument of the scaling function $g$ fixed to
$\delta \theta L^\phi=94.8$ and
the crossover exponent $\phi=1.6$
for various system sizes,
($+$) $L=16$,
($\times$) $18$,
and
($*$) $20$;
see the crossover-scaling formula
(\ref{crossover_scaling_formula}).
The crossover-scaled data fall into a scaling curve.
Particularly,
the data, 
	($\times$) $L=18$ and 
	($*$) $20$,
are about to overlap each other, suggesting that the
proposition $\phi=1.6$ is an optimal one.
}
\label{figure6}       
\end{figure}

\begin{figure}
\includegraphics{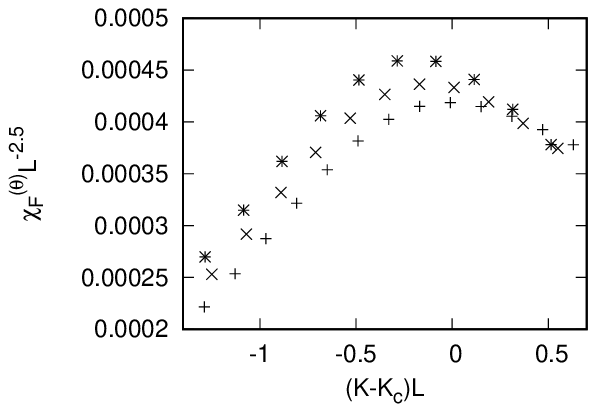}
\caption{
The crossover-scaling plot,
$(K-K_c(\theta))L$-$\chi_F^{(\theta)}L^{-2.5}$,
is presented 
with the
second argument of the scaling function $g$ fixed to
$\delta \theta L^\phi=233$ and
the crossover exponent $\phi=1.9$;
see the crossover-scaling formula
(\ref{crossover_scaling_formula}).
The symbols,
($+$),
($\times$),
and
($*$),
are the same as those of Fig. \ref{figure6}.
The left-side slope gets resolved as for large $\phi$.
}
\label{figure7}       
\end{figure}

\begin{figure}
\includegraphics{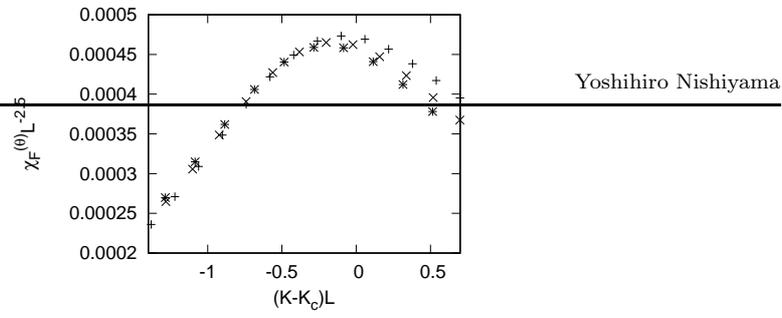}
\caption{
The crossover-scaling plot,
$(K-K_c(\theta))L$-$\chi_F^{(\theta)}L^{-2.5}$,
is presented 
with the
second argument of the scaling function $g$ fixed to
$\delta \theta L^\phi=38.6$ and
the crossover exponent $\phi=1.3$;
see the crossover-scaling formula
(\ref{crossover_scaling_formula}).
The symbols,
($+$),
($\times$),
and
($*$),
are the same as those of Fig. \ref{figure6}.
The right-side slope splits off as for small $\phi$.
}
\label{figure8}       
\end{figure}

In Fig. \ref{figure6},
we
present the crossover-scaling plot,
$(K-K_c(\theta))L$-
$\chi_F^{(\theta)}L^{-2.5}$,
for various system sizes, 
($+$) $L=16$,
($\times$) $18$, and
($*$) $20$.
Here, the second argument of the scaling function
$g$ is fixed to
a constant value,
$\delta \theta L^{\phi/\dot{\nu}} =94.8$, with an optimal crossover exponent
$\phi=1.6$,
and
the critical point $K_c(\theta)$ was determined with the same
scheme as that of Sec. \ref{section2_1}.
The crossover-scaled data in Fig. \ref{figure6}
collapse into a scaling curve;
particularly, the data, ($\times$) $L=18$ and ($*$) $20$,
are about to overlap each other, showing a tendency to the convergence as $L\to\infty$.
Likewise, in Fig. \ref{figure7} and \ref{figure8},
we present the crossover-scaling plot,
$(K-K_c(\theta))L$-$\chi_F^{(\theta)}L^{-2.5}$,
with the crossover exponent,
$\phi=1.9$ and $1.3$, respectively;
the symbols are the same as those of Fig. \ref{figure6}.
Here, the second argument
of the scaling function $g$
is set to 
$\delta \theta L^{\phi/\dot{\nu}} =233$ and
$38.6$ in the respective analyses.
In the former (latter) scaling plot,
the left- (right-) side slope starts to split off,
indicating that 
even larger (smaller) parameter $\phi$ leads a scatter of the scaled data.
Hence, considering that these parameters set the tolerable bounds,
we estimate the crossover exponent
as
\begin{equation}
	\label{crossover_exponent}
\phi=1.6(3)
.
\end{equation}

This is a good position to
address a number of remarks.
First,
the underlying physics behind the crossover-scaling plot, Fig. \ref{figure6},
differs from that of the fixed-$\theta$ scaling plot, Fig. \ref{figure5}.
Actually, the former scaling dimension $\dot{x}=5/2$ is much larger than
the latter $x=1$, and hence, the data collapse of the crossover-scaling plot is
by no means accidental.
Second, the honeycomb-lattice Ising antiferromagnet enjoys the duality theory \cite{Imaoka96} 
so as to fix the critical indices such as $\dot{\nu}=1$ and $\dot{\alpha}_F^{(\theta)}/\dot{\nu}=5/2$
\cite{Horiguchi92}.
Hence, it is anticipated that the crossover exponent $\phi=1.6(3)$ 
[Eq. (\ref{crossover_exponent})] 
reflects the peculiarities of the honeycomb-lattice structure.
Actually, the estimate $\phi=1.6(3)$ 
[Eq. (\ref{crossover_exponent})]
differs from the mean-field value $\phi=1/2$ \cite{Azcoiti11}, 
whereas
it is slightly suppressed as compared to 
the square-lattice case, $\phi=2.0(4)$ \cite{Nishiyama20}.
Because the magnetic-susceptibility index for the honeycomb lattice $\dot{\gamma}_{af}=3/2$ 
\cite{Horiguchi92}
is substantially smaller than that of the square lattice $5/2$ \cite{Matveev95},
it is reasonable that the multi-criticality depends on each lattice structure undertaken.
Last, we mention a candidate for the quantifier other than the fidelity susceptibility.
So far,
the correlation length has played a significant role in the finite-size-scaling analyses.
Actually,
it 
is accessible via the diagonalization method
\cite{Forcrand18}, provided that the second-largest eigenvalue of the transfer matrix is at hand.
The correlation length has an advantage in that it has a fixed scaling dimension $\sim L^1$
{\it a priori}.
However, the second-largest eigenvalue
is computationally demanding particularly for the non-hermitian transfer matrix,
and this scheme was not accepted here.

\section{\label{section3}
Summary and discussions}

The honeycomb-lattice Ising antiferromagnet (\ref{Hamiltonian})
under the imaginary magnetic field $H=i\theta T/2$
was investigated with 
the transfer-matrix method \cite{Forcrand18}.
As a probe to detect the phase transition,
we utilized
the extended version \cite{Schwandt09,Sirker10}
of the fidelity (\ref{extended_fidelity}),
which makes sense even for such a non-symmetric 
complex-valued transfer matrix.

As a demonstration, 
we investigated the order-disorder phase transition for $\theta=2.22$
with the fidelity susceptibility
$\chi_F^{(\theta)}$ (\ref{fidelity_susceptibility}).
Our result $K_c|_{\theta=2.22}=0.484(4)$ 
[Eq. (\ref{critical_point})]
is comparable to that of the partition-function-zeros method,
$K_c|_{\theta=2.22}=0.458 \dots$
\cite{Kim10},
indicating that the probe
$\chi_F^{(\theta)}$ detects the phase transition sensitively
even in the presence of the imaginary
magnetic field.
Furthermore, we estimated $\chi_F^{(\theta)}$'s scaling dimension as $\alpha_F^{(\theta)}/\nu=0.97(3)$
[Eq. (\ref{critical_exponent})].
Through resorting to the scaling relation (\ref{scaling_relation}),
we estimated magnetic-susceptibility's scaling dimension as $\gamma_{af}/\nu=-0.03(3)$ 
[Eq. (\ref{critical_exponent_gamma})].
This result indicates that the criticality belongs to the
two-dimensional-Ising universality class,
$\gamma_{af}=0$ (logarithmic) \cite{Fisher60,Kaufman87}.
We then turn to the analysis of the end-point singularity of the phase boundary
toward 
$\delta  \theta(=\pi-\theta) \to 0$.
With $\delta \theta$ scaled carefully,
the $\chi_F^{(\theta)}$ data are cast into the crossover-scaling
formula (\ref{crossover_scaling_formula}) \cite{Riedel69,Pfeuty74}.
Thereby, we estimated
the crossover exponent as $\phi=1.6(3)$ 
[Eq. (\ref{crossover_exponent})].
This result
differs
from the mean-field value $\phi=1/2$ \cite{Azcoiti11},
whereas it is slightly suppressed as compared to that of the square-lattice model,
$\phi=2.0(4)$ \cite{Nishiyama20}.
It would be intriguing that 
the lattice structure renders subtle influences as to the
end-point singularity.
Actually, the honeycomb-lattice antiferromagnet at $\theta=\pi$ 
is under the reign of
the duality theory
\cite{Imaoka96,Suzuki90,Lin88,Matveev96}, 
and it is anticipated that the end-point singularity
reflect its peculiar characters.

According to Refs. \cite{Suzuki90,Matveev08,Sarkanch18},
even in the ferromagnetic side $K<0$, 
there should occur a singularity for generic values of $0<\theta<\pi$;
a notable point is that the transition is {\em not} the ordinary
order-disorder phase transition \cite{Suzuki90}.
According to 
the partition-function-zeros survey \cite{Matveev08}, the transition point $|K_c|$
should locate around $\exp(-4|K_c|) \approx 0.8(\ne 1)$ 
at $\theta=\pi/2$ as for the square lattice.
Because the fidelity-susceptibility-mediated analysis does not require any {\it a priori}
settings as to the order parameter,
it would 
provide valuable information even for such an exotic singularity.
This problem is left for the future study.

\section*{Acknowledgment}
This work was supported by a Grant-in-Aid
for Scientific Research (C)
from Japan Society for the Promotion of Science
(Grant No. 
20K03767).

{\bf Author contribution statement}

Y.N. conceived the presented idea,
and performed the numerical simulations.
He analyzed the numerical data, and wrote up
the manuscript.

\begin{table}
\caption{
A comparison is made between the 
partition-function-zeros analysis
\cite{Kim10} and ours.
From an accumulation of zeros toward 
the unit circle
$x=e^{i\theta}$
as shown in Fig. 2 (d) of Ref. \cite{Kim10},
we read off
the critical point
$K_c=0.458
\dots(=-\ln 0.4/2)$
at $\theta=2.22\dots$;
see text for details.
Our transfer-matrix method with the aid of the 
extended
fidelity susceptibility
(\ref{fidelity_susceptibility})
appears to support this elaborated pioneering study.
}
\label{table1}       
\begin{tabular}{llll}
\hline\noalign{\smallskip}
 method & quantifier &  $K_c|_{\theta=2.22}$ \\
\noalign{\smallskip}\hline\noalign{\smallskip}
partition-function zeros \cite{Kim10} & accumulation of zeros &  $0.458\dots$ \\  
transfer matrix (present work) & fidelity susceptibility &   $0.484(4)$  \\
%
\noalign{\smallskip}\hline
\end{tabular}
\end{table}

%
%




\end{document}